\documentclass[twocolumn,showpacs,preprintnumbers,amsmath,amssymb]{revtex4}
\def\a {\epsilon}

\def\F  {{\cal F}}
\def\O  {{\cal O}}
\def\R  {{\cal R}}
\def\bc {$B_c^{(*)}$ }

\def\ktf {$k_t$-factorization }

\def\cpc#1#2#3  {{Chin.\ Phys.\ C }          {\bf#1}, #2 (#3)}
\def\err#1#2#3  {{\it Erratum }              {\bf#1}, #2 (#3)}
\def\epjc#1#2#3 {{Eur.\ Phys.\ J.\ C }       {\bf#1}, #2 (#3)}
\def\dum#1#2#3  {{~}                         {\bf#1}, #2 (#3)}
\def\ib#1#2#3   {{\it ibid. }                {\bf#1}, #2 (#3)}
\def\jcp#1#2#3  {{J.\ Comp.\ Phys.\ }        {\bf#1}, #2 (#3)}
\def\jhep#1#2#3 {{JHEP }                     {\bf#1}, #2 (#3)}
\def\ijmp#1#2#3 {{Int.\ J.\ Mod.\ Phys.\ }   {\bf#1}, #2 (#3)}
\def\jpg#1#2#3  {{J.\ Phys.\ G }             {\bf#1}, #2 (#3)}
\def\mpl#1#2#3  {{Mod.\ Phys.\ Lett.\ }      {\bf#1}, #2 (#3)}
\def\ncim#1#2#3 {{Nuovo Cimento }            {\bf#1}, #2 (#3)}
\def\np#1#2#3   {{Nucl.\ Phys.\ }            {\bf#1}, #2 (#3)}
\def\npb#1#2#3  {{Nucl.\ Phys.\ B }          {\bf#1}, #2 (#3)}
\def\pan#1#2#3  {{Phys.\ At.\ Nuclei }       {\bf#1}, #2 (#3)}
\def\plb#1#2#3  {{Phys.\ Lett.\ B }          {\bf#1}, #2 (#3)}
\def\prep#1#2#3 {{Phys.\ Rep.\ }             {\bf#1}, #2 (#3)}
\def\prd#1#2#3  {{Phys.\ Rev.\ D }           {\bf#1}, #2 (#3)}
\def\prl#1#2#3  {{Phys.\ Rev.\ Lett.\ }      {\bf#1}, #2 (#3)}
\def\ptp#1#2#3  {{Prog.\ Theor.\ Phys.\ }    {\bf#1}, #2 (#3)}
\def\ps#1#2#3   {{Physica Scripta }          {\bf#1}, #2 (#3)}
\def\rmp#1#2#3  {{Rev.\ Mod.\ Phys.\ }       {\bf#1}, #2 (#3)}
\def\rpp#1#2#3  {{Rep.\ Prog.\ Phys.\ }      {\bf#1}, #2 (#3)}
\def\sa#1#2#3   {{Sci.\ Acta }               {\bf#1}, #2 (#3)}
\def\sjnp#1#2#3 {{Sov.\ J.\ Nucl.\ Phys.\ }  {\bf#1}, #2 (#3)}
\def\spj#1#2#3  {{Sov.\ Phys.\ JETP }        {\bf#1}, #2 (#3)}
\def\spjl#1#2#3 {{Sov.\ JETP Lett.\ }        {\bf#1}, #2 (#3)}
\def\spu#1#2#3  {{Sov.\ Phys.-Usp.\ }        {\bf#1}, #2 (#3)}
\def\yaf#1#2#3  {{Yad.\ Fiz.\ }              {\bf#1}, #2 (#3)}
\def\zp#1#2#3   {{Zeit.\ Phys.\ }            {\bf#1}, #2 (#3)}
\def\zpa#1#2#3  {{Z.\ Phys.\ A }             {\bf#1}, #2 (#3)}
\def\zpc#1#2#3  {{Z.\ Phys.\ C }             {\bf#1}, #2 (#3)}
\def\et {{\it et al.}}
\newcommand{\n}{\nonumber \\}

\usepackage{graphics}
\usepackage{amsmath}
\usepackage{epsfig}

\begin{document}
\draft
\title{First estimates of the $B_c$ wave function \\
  from the data on the $B_c$ production cross section }
\author{S.P.\ Baranov}
\email{baranov@sci.lebedev.ru}
\affiliation{P.N. Lebedev Institute of Physics, 
              Lenin Avenue 53, 119991 Moscow, Russia}
\author{A.V.\ Lipatov}
\email{lipatov@theory.sinp.msu.ru}
\affiliation{Skobeltsyn Institute of Nuclear Physics,
      Lomonosov Moscow State University, 119991 Moscow, Russia}
\affiliation{Joint Institute for Nuclear Research, 141980 Dubna, Moscow Region, Russia}
%

\date{\today}
\begin{abstract}
In the framework of perturbative QCD and nonrelativistic bound state
formalism, we calculate the production of $B_c$ and $B_c^*$ mesons at the 
conditions of the CDF and LHCb experiments. We derive first estimations 
for the $B_c$ wave function from a comparison with the available data. 
\end{abstract}
\maketitle

\section{INTRODUCTION}

The family of \bc mesons is an interesting though poorly explored part
of the quarkonium world. 
Although some properties of these mesons may look apparently different 
from the ones of the hidden-flavor onium states, their inner structure 
must be similar and driven by the same physics.
Studying the \bc properties is important on its own and can provide an
additional cross check of the exploited theoretical models.

The flavor composition of \bc mesons excludes the convenient strong 
and electromagnetic decays channels that could be used as a prompt measure 
of the nonrelativistic wave function. Instead, we will try to obtain an 
estimate of this essential parameter via considering the production process.
We will rely on the data collected by the CDF Collaboration at the Fermilab 
Tevatron at 1.8 TeV \cite{CDF_1998} and 1.96 TeV \cite{CDF_2016} and by the 
LHCb Collaboration at CERN LHC at 7 TeV \cite{LHCb_2012} and 8 TeV 
\cite{LHCb_2015}.

\section{THEORETICAL FRAMEWORK}

In the theory, the production of \bc mesons at the LHCb conditions is 
dominated by the $\O{(\alpha_s)^4}$ partonic subprocess
\begin{equation}
      g+g\to B_c^{(*)} + b + \bar{c}, \label{gluglu}
\end{equation}
where \bc may denote either pseudoscalar $B_c$ (spin=0) or vector $B_c^*$ 
(spin=1) bound state of the charm and beauty quarks.
The evaluation of the relevant 36 Feynman diagrams is straightforward
and is described in every detail in Ref. \cite{myBc}. The only innovation 
made in the present calculation is in using the \ktf approach.
The advantage of the latter comes from the ease of including the initial 
state radiation corrections that are efficiently taken into account in the 
form of the evolution of gluon densities. Then, in accordance with the 
\ktf prescription \cite{GLR83}, the initial gluon spin density matrix is
taken in the form
$\overline{\epsilon_g^{\mu}\epsilon_g^{*\nu}}=k_T^\mu k_T^\nu/|k_T|^2,$ 
where $k_T$ is the component of the gluon momentum perpendicular to the 
beam axis. In the limit when $k_T\to 0,$ this expression converges to the 
ordinary $\overline{\epsilon_g^{\mu}\epsilon_g^{*\nu}}=-g^{\mu\nu}/2$, and we
recover the results of collinear approach \cite{Chen,Ruckl,Likhoded}. 
This work is the first calculation of the \bc hadronic production with 
$k_t$-factorization.

The perturbative part of our calculation is performed according to the
formula 
\begin{eqnarray}
&& \!\!\!\!\!\!\!\! d\sigma(pp\to B_c b\bar{c}X)= \n 
&& \frac{\alpha_s^4}{12\,\hat{s}^2}\,     |{\R}(0)|^2 
   \frac{1}{4}\sum_{\mbox{{\tiny spins}}}\;\frac{1}{64}\sum_{\mbox{{\tiny colors}}}
   |{\cal M}(gg\to B_c b\bar{c})|^2 \n 
&& \times {\F}_g(x_1,k_{1T}^2,\mu^2)\;{\F}_g(x_2,k_{2T}^2,\mu^2)\;
   dk_{1T}^2\,dk_{2T}^2 \n
&& \times dp_{B_cT}^2\,dp_{cT}^2\,dy_{B_c}\,dy_b\,dy_{\bar{c}}\,
\frac{d\phi_1}{2\pi} \frac{d\phi_2}{2\pi}
\frac{d\phi_{B_c}}{2\pi} \frac{d\phi_c}{2\pi}, \label{lips}
\end{eqnarray}
and we use the JH'2013 (set 2) \cite{JH2013} parametrization 
for the transverse momentum dependent (TMD, or unintegrated) gluon distribution ${\F}_g(x_i,k_{iT}^2,\mu^2)$.

The absolute normalization of the cross section depends on a singe
non-perturbative parameter: the radial (color singlet) wave function at 
the origin of the coordinate space $|{\R}(0)|^2$ \cite{Chang,Jones,Baier}.
The so called color octet contributions are not important as they are
suppressed by the relative velocity counting rules. Note by the way that
the gluon fragmentation mechanism (known to dominate the production of
$J/\psi$ mesons at high transverse momenta) is not applicable to our case.

To perform a comparison with the data (see below) we also need to 
calculate the production of the ordinary $B^+$ mesons. Again, we do that 
in the \ktf approach with the same gluon density \cite{JH2013} as above,
and with Peterson \cite{Peterson} fragmentation function with $\a=0.0126$
for the formation of $B^+$ mesons from $b$-quarks. The consistency of this
setting was shown in a previous publication \cite{bb}.

\section{NUMERICAL RESULTS AND DISCUSSION}

The data we wish to compare with are presented in the form of the ratio
of the \bc to $B^+$ production cross sections times the relevant branching
fractions. All these results accumulate the statistics from both $B_c$ and 
$B_c^*$ mesons and include also their charge conjugate states.

Ref. \cite{CDF_1998} reports for the fiducial phase space defined as
$p_T^{B_c}>6$ GeV, $p_T^{B^+}>6$ GeV, $|y^{B_c}|<1$, $|y^{B^+}|<1$:
\begin{equation}
\frac{\sigma(B_c)\,Br(B_c\to J/\psi\,l\nu)}{\sigma(B^+)\,Br(B^+\to J/\psi\,K)}
= 0.132\pm \begin{smallmatrix}0.061\\ 0052\end{smallmatrix}.
\end{equation}
Hereafter, in the experimental references $B_c$ will denote a combined
sample of $B_c$ and $B_c^*$ mesons.
Within the specified kinematic cuts, we obtain from Eq.(2):
\begin{eqnarray}
\sigma^{theor}(B_c)  &=& |{\R}(0)|^2\cdot 0.247\mbox{~nb/GeV}^3 ,\n
\sigma^{theor}(B_c^*)&=& |{\R}(0)|^2\cdot 0.516\mbox{~nb/GeV}^3, \n
\sigma^{theor}(B_c{+}B_c^*)&=& |{\R}(0)|^2\cdot 0.763\mbox{~nb/GeV}^3.
\end{eqnarray}
We also have for the production of $B^+$ mesons
\begin{equation}
\sigma^{theor}(B^+)\,Br(B^+\to J/\psi\,K^+)=7.13\mbox{~nb},
\end{equation}
where we have used the decay branching fraction value
\begin{equation}
Br(B^+\to J/\psi\,K^+)=1.026\cdot 10^{-3}
\end{equation}
taken from the Particle Data book \cite{PDG}.

Ref. \cite{CDF_2016} reports for
$p_T^{B_c}>6$ GeV, $p_T^{B^+}>6$ GeV, $|y^{B_c}|<0.6$, and $|y^{B^+}|<0.6$:
\begin{equation}
\frac{\sigma(B_c)\,Br(B_c\to J/\psi\,l\nu)}{\sigma(B^+)\,Br(B^+\to J/\psi\,K)}
 = 0.211\pm \begin{smallmatrix}0.024\\ 0.023\end{smallmatrix}
\end{equation}
Within the above cuts, we obtain 
\begin{eqnarray}
\sigma^{theor}(B_c)  &=& |{\R}(0)|^2\cdot 0.177\mbox{~nb/GeV}^3, \n
\sigma^{theor}(B_c^*)&=& |{\R}(0)|^2\cdot 0.3646\mbox{~nb/GeV}^3, \n
\sigma^{theor}(B_c{+}B_c^*)&=& |{\R}(0)|^2\cdot 0.541\mbox{~nb/GeV}^3,
\end{eqnarray}
and 
\begin{equation}
\sigma^{theor}(B^+)\,Br(B^+\to J/\psi\,K^+)=4.99\mbox{~nb}.
\end{equation}

Ref. \cite{LHCb_2012} reports for $p_T^{B_c}>4$ GeV, $p_T^{B^+}>4$ GeV, 
$2.0<|y^{B_c}|<4.5$, and $2.0<|y^{B^+}|<4.5$:
\begin{equation}
\frac{\sigma(B_c)\,Br(B_c\to J/\psi\,\pi^+)}{\sigma(B^+)\,Br(B^+\to J/\psi\,K)}
= 0.0061\pm 0.0012.
\end{equation}
Under these conditions, we have:
\begin{eqnarray}
\sigma^{theor}(B_c)  &=& |{\R}(0)|^2\cdot 2.37\mbox{~nb/GeV}^3, \n
\sigma^{theor}(B_c^*)&=& |{\R}(0)|^2\cdot 3.03\mbox{~nb/GeV}^3, \n
\sigma^{theor}(B_c{+}B_c^*)&=& |{\R}(0)|^2\cdot 5.40\mbox{~nb/GeV}^3,
\end{eqnarray}
and
\begin{equation}
\sigma^{theor}(B^+)\,Br(B^+\to J/\psi\,K^+)=27.3\mbox{~nb}.
\end{equation}

\begin{figure}[b]
\includegraphics[width=4.25cm]{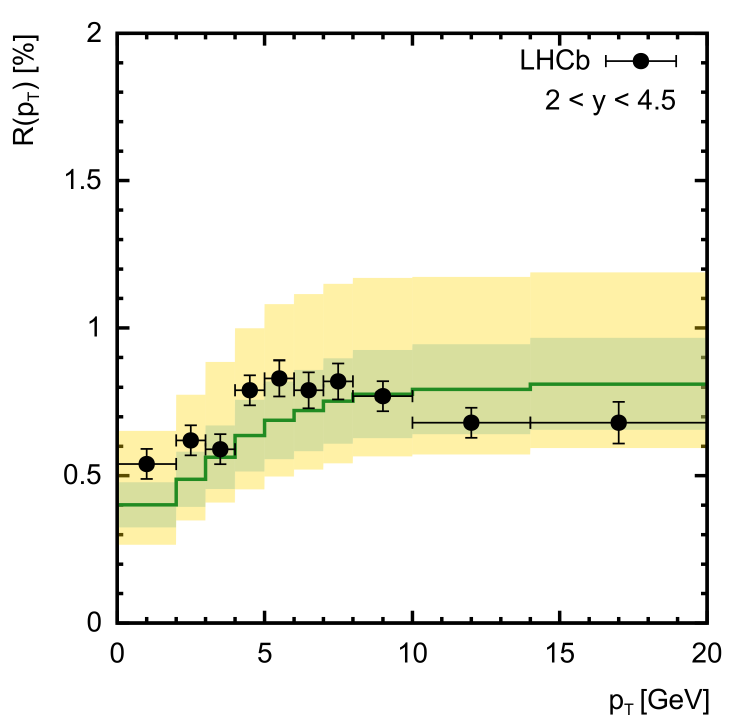}
\includegraphics[width=4.25cm]{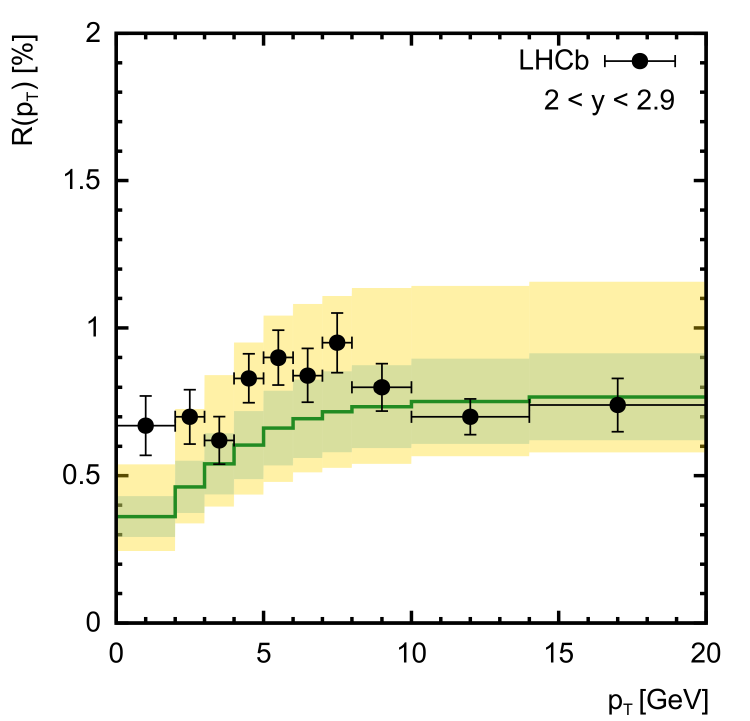}
\includegraphics[width=4.25cm]{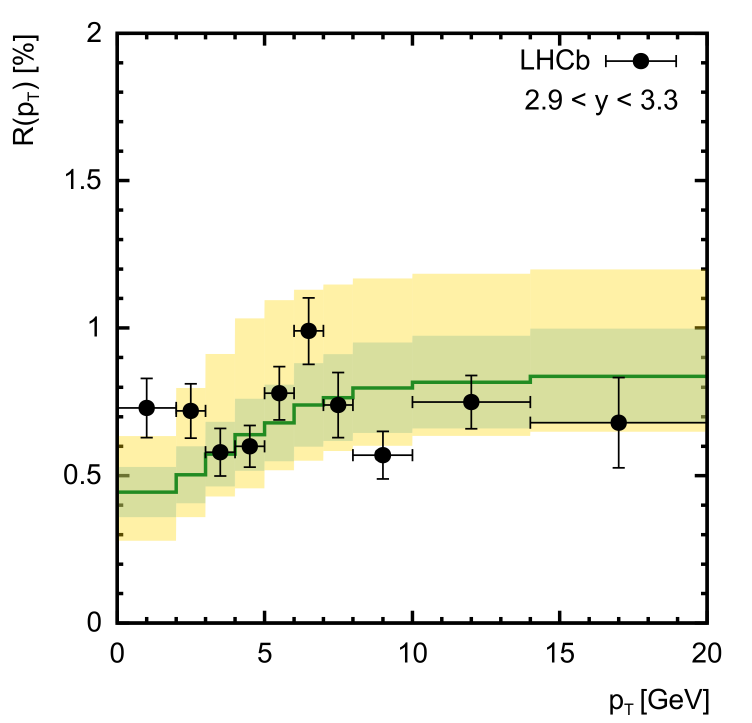}
\includegraphics[width=4.25cm]{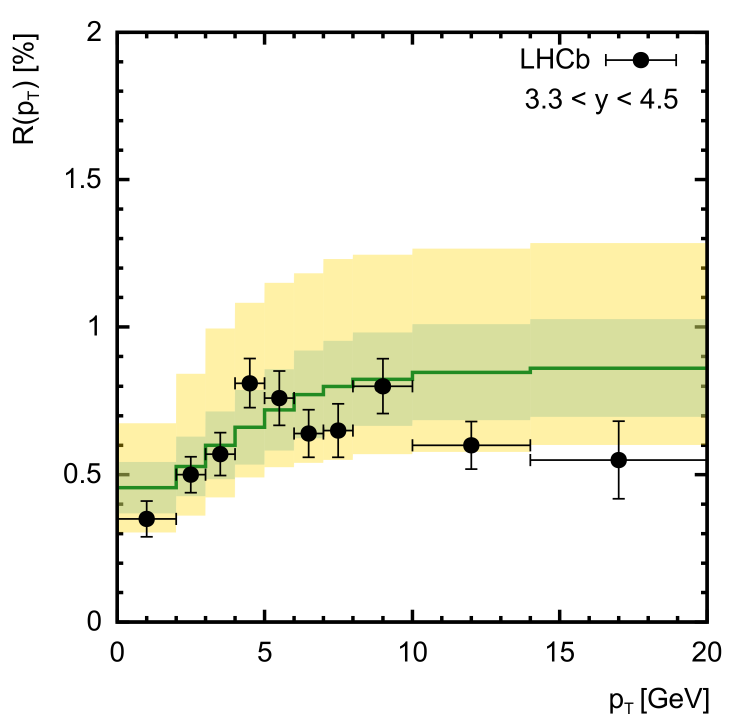}
\caption{%
The ratio of the \bc to $B^+$ production cross sections (13) as a function
of the transverse momentum for 
different rapidity intervals.
Grey band indicates the uncertainty in the determination of the \bc
radial wave function; yellow band represents uncertainties coming from
the renormalization scale. 
Experimental points
are from LHCb \cite{LHCb_2015}.
}
\end{figure}

Finally, Ref. \cite{LHCb_2015} reports for 
$p_T^{B_c}<20$ GeV, $p_T^{B^+}<20$ GeV,
$2.0<|y^{B_c}|<4.5$, and $2.0<|y^{B^+}|<4.5$:
\begin{equation}
\frac{\sigma(B_c)\,Br(B_c\to J/\psi\,\pi^+)}{\sigma(B^+)\,Br(B^+\to J/\psi\,K)}
= 0.0068\pm 0.0002;
\end{equation}\\
and our predictions read:
\begin{eqnarray}
\sigma^{theor}(B_c)  &=& |{\R}(0)|^2\cdot 4.92\mbox{~nb/GeV}^3, \n
\sigma^{theor}(B_c^*)&=& |{\R}(0)|^2\cdot 5.63\mbox{~nb/GeV}^3, \n
\sigma^{theor}(B_c{+}B_c^*)&=& |{\R}(0)|^2\cdot 10.55\mbox{~nb/GeV}^3,
\end{eqnarray}
and
\begin{equation}
\sigma^{theor}(B^+)\,Br(B^+\to J/\psi\,K^+)=65.33\mbox{~nb}.
\end{equation}

The above data have to be combined with the experimentally measured 
\cite{LHCb_2014} ratio of the branching fractions
\begin{equation}
Br(B_c\to J/\psi\,\pi^+)/Br(B_c\to J/\psi\,\mu\nu)=0.047
\end{equation}
and with the theoretically calculated \cite{Qiao} decay branching fraction
\begin{equation}
Br(B_c\to J/\psi\,\pi^+)=0.0033. 
\end{equation}
The original \cite{Qiao} prediction of 0.0029 was corrected 
\cite{LHCb_2015} to 0.0033 for the latest measurement of the $B_c$ lifetime.

Making the necessary substitutions and comparing Eqs. (3), (7), (10),
and (13) with theoretical predictions we deduce the following estimations
for the radial wave function:
\begin{eqnarray}
|{\R}(0)|^2 = 4.40 \pm 2.00 \mbox{~GeV}^3 && \mbox{Ref. \cite{CDF_1998} }\n
|{\R}(0)|^2 = 6.91 \pm 0.08 \mbox{~GeV}^3 && \mbox{Ref. \cite{CDF_2016} }\n
|{\R}(0)|^2 = 5.15 \pm 0.10 \mbox{~GeV}^3 && \mbox{Ref. \cite{LHCb_2012} }\n
|{\R}(0)|^2 = 7.05 \pm 0.20 \mbox{~GeV}^3 && \mbox{Ref. \cite{LHCb_2015} }
\end{eqnarray}
These can be summarised in a mean-square average value
\begin{equation}
|{\R}(0)|^2 = 5.88 \mbox{~GeV}^3
\end{equation}
with an error of $\pm 0.64$ GeV$^3$ and $\pm 1.07$ GeV$^3$ at the
60\% and 80\% confidence level, respectively. 

We conclude our analysis with showing the ratio (13) in the differential
form, as a function of the transverse momentum for several rapidity
intervals (see Fig.~1). The calculations and the data \cite{LHCb_2015}
are in good agreement in shape, thus indicating that the hard scattering
partonic subprocesses are calculated correctly. The choice of the TMD gluon
parametrization is unimportant since the gluon distributions cancel out
in the ratio. The sensitivity to the renormalization scale is high, as a 
reflection of the high power of $\alpha_S(\mu_R^2)$ in the key subprocess (1).
The central values of the cross sections correspond to the conventional
choice $\mu_R^2=p_{B_c T}^2+m_{B_c}^2;$ the theoretical uncertainty band
(yellow area in Fig. 1) is obtained by varying $\mu_R$ around its default
value by a factor of 2.

Our extracted values of $|{\R}(0)|^2$ are higher 
than the predictions \cite{Quigg} of potential models
which range from 1.508~GeV$^3$ for the logarithmic potential \cite{pot10}, 
through 1.642~GeV$^3$ and 1.710~GeV$^3$ for the Buchmuller-Tye \cite{pot8} 
and power low \cite{pot9} potentials up to 3.102~GeV$^3$ for the Cornell 
potential \cite{pot4}. This may be taken as an evidence of the importance 
of radiative corrections (the latter are known to be large for $J/\psi$ 
mesons). Another possible interpretation may guess that the conventional 
choice of $\mu_R$ somehow overestimates the momentum transfer in the hard 
process. Any way, the agreement between the theory and the data is rather 
satisfactory and shows no fundamental problems in describing the data.



\acknowledgments

We would like to thank H.~Jung and M.A.~Malyshev for very useful 
discussions and important remarks.
This work was supported by the DESY Directorate in the framework of 
Moscow-DESY project on Monte-Carlo implementations for HERA-LHC.

%

\end{document}